# Photocurrent in Bismuth Junctions with Graphene

*Tito E. Huber,[1,a] Tina Brower,[1] Scott. D. Johnson,[1] John H. Belk[3] and Jeff H. Hunt[4]*

*[1]Howard University, 500 College St. NW, Washington, DC 20059. USA*

*[3] The University of Missouri, MO 63134, USA*

*[4] The Boeing Company. 900 N Sepulveda Blvd, El Segundo, CA 90245. USA*

Abstract

We report on a room-temperature photodetector utilizing semimetal bismuth nanowire arrays coupled with graphene. The structural flatness and high electron mobility of graphene exhibit great potential for future generations of electronic materials. Bismuth nanowire arrays coupled with graphene have strong absorption because of light trapping. Bismuth (Bi), as a semimetal, complements graphene's gapless and mobility characteristics. Bi also features a long screening length (4 nm) comparable to semiconductors. Raman spectroscopy is used to characterize the charge transfer between graphene and Bi. The analyzed spectrum includes the visible (350 nm) through the near infrared (980 nm) and well into the telecommunication band (1550 nm). Because of photocarrier pair generation and transfer at the interface, the photocurrent generated by the interface built-in field is robust without sacrificing the detection spectrum. We observe a power-law frequency dependence of the photoresponse over three order of magnitude of excitation frequency consistent with a wide distribution of diffusion times. These key properties may enable application of the junctions of graphene with Bi devices in detector and light harvesting applications.

Key points: Graphene. Semimetal. Light trapping. Raman Photovoltaic. Photothermoelectric.

Graphene, the novel two-dimensional (2D) material with an atomic "honeycomb" lattice, has drawn strong interest due to its extraordinary electrical properties.[1] The high charge carrier mobility in graphene[2] makes it a promising candidate for the new generation of photonic and optoelectronic devices. Photodetectors based on single-layer graphene (SLG) demonstrate broad detecting spectral widths[3] related to the gapless nature of its conduction band. SLG has limited absorption (2 %), inhibiting efficient photo carrier separation and accumulation.[4,5] Many efforts focus on modifying graphene so that it has a bandgap while at the same time preserving the high electronic mobility characteristic of graphene[6,7]. Another area of investigation is the integration of graphene with metals and semiconductors in bulk form and 2D materials, such as transition metal chalcogenides (TMDCs)[8,9,10] or Bi(2)Te(3).[11] However, the gap in the semiconductors can limit the application of the hybrids in the infrared. Bismuth has not been explored extensively in interfaces with graphene although it has intriguing properties. Like graphene, bismuth is a semimetal. Bi has a distinct Fermi surface that features small electron pockets centered at the three L points of the Brillouin zone and also a small hole pocket at the T point. Bi electron $n$ and hole $p$ density ($n=p$) of $10^{18}$ cm$^{-3}$ is low, more comparable to that found in lightly doped semiconductors than metals, where the density is ~ $10^{23}$ cm$^{-3}$. Bismuth Thomas-Fermi screening length is exceptionally long, approximately 4 nm.[12] In contrast, $\lambda$ is only a fraction of a nanometer for metals. The long screening length of semimetals is comparable to that of lightly doped semiconductors. However, semiconductors feature a Schotky barrier, that is absent in metals.

Free carrier absorption processes in bismuth, that dissipate photon energy without creating electron-hole pairs, can be abated by shaping the material in nanowires where the electrons are restricted to move in paths oriented perpendicular to the light polarization. Therefore, the optical absorption can be dominated by electron-hole pair processes. The indirect L-T transition that starts at 1000 cm$^{-1}$ and the direct L point transition, that starts at 2000 cm$^{-1}$, have been observed in the optical absorption in nanowires in the infrared.[13-15] Control of optical free carrier absorption in nanowires is a key part of the design of our device. Recently, Yao et al reported the observation of broadband photoconductivity of bismuth films under ultraviolet to near infrared (1550 nm) light.[16] This observation is attributed to electron-hole pairs. On light absorption, the generated electron-hole pairs would normally recombine in a short time scale. When an external field is applied, the pairs are separated and a photocurrent is generated. The same would happen in the presence of an internal or built-in field[17-18] however this has never been observed for bismuth. Such fields are shown to be formed near the graphene-metal interfaces so it can be assumed that the same happens in graphene-semimetal interfaces. This is the phenomenon that we want to explore in bismuth-graphene junctions. Built-in fields manifest themselves as charge transfer to graphene and we present a study of the interface via Raman spectroscopy that allows us to evaluate charge transfer.

Electrically, the interfaces of graphene to conductors can be Ohmic or rectifying and they may even feature a Schottky barrier like semiconductor-graphene interfaces.[19] Here we present an electrical characterization of the interface between graphene and bismuth using current-voltage (I-V) techniques.

We show that the contact is Ohmic in our case. Such case is characteristic of metal-graphene interfaces rather than semiconductor-graphene interfaces. Finally, we assembled a device to measure the photoresponse. We employed various wavenumbers from the visible to the infrared.

Like metals, the semimetal bismuth features high optical reflectivity. Our approach to the high reflectivity of bismuth is light trapping because this is natural for a nanowire that presents a high conductance path to the photocurrent. Light trapping consists of broadband optical absorption by nanowires oriented in the optical incident direction. Light trapping can be traced to the optical properties of individual nanowires.[20-30] Arrays of independent optical antennas have an optical response that is polarization-independent and angle-insensitive. For these sparse nanowire arrays, when the wire radius is uniform, strong absorption is observed over a relatively narrow spectral region in which end-mediated coupling into guided modes is favorable.[20] In high density arrays, there is strong inter-nanowire interaction and the absorption features broaden.[21-30] Array geometry, nanowire shape, and order have previously been shown, both experimentally and theoretically, to control the spectral position and breadth of the absorbing region. The non-uniformity of the nanowire arrays also causes broadening. Light trapping is exploited in photovoltaic cells based on semiconductor nanowire arrays capped with graphene; a promising candidate for solar energy harvesting. Well-known examples are nanowire arrays of silicon, ZnO and GaN coated with graphene.[31-35]. Here, we present an investigation of Bi nanowire arrays capped with graphene, a case that has not been explored. The concept of built-in field is important in photoemission. In the case of our graphene hybrids, the built-in field is associated with the work functions of bismuth and graphene and the potential barriers in the contact. Khomyakov *et al* presents a study of the absorption of graphene on metal substrates based on first-principles calculations.[36] Guided by this work, we consider that the modifications of SLG that we observe via Raman can be associated with contact potential barriers and changes in graphene Fermi level.[37-42] We will also discuss hot carriers,[43-45] photo-thermoionic,[46] and thermoelectric response.[47-48]

RESULTS AND DISCUSSION

Figure 1a illustrates the configuration of the photodetector device based on the graphene−NWA junctions. We employ hybrid heterostructures composed of bismuth nanowire arrays which are capped with SLG. Dense arrays of 200-nm nanowires have been prepared by a fabrication technique consisting of the pressure injection of an alumina template with molten Bi, a method that can be successfully employed with 100-µm thick templates with pore diameters in the range of 2 to 200 nm. For the experiments reported here, we fabricated 200-nm wires. The scanning electron microscope (SEM) image is shown in Figure 1b. The sample is 1.5 mm × 2 mm. Its I-V is linear at room temperature (the only temperature that was investigated) and its resistance is 90 Ohms. Fig 1c is an optical image of the interface. The surface appears black because of light trapping. The optical field penetrates in the structure at least a wavelength. This is revealed by the polarization of reflected light when the direction of incidence is off-normal.   The nanowire arrays electronic properties including electronic transport, magnetotransport and thermopower, are studied in separate experiments.[48] The SLG is fabricated by chemical vapor deposition (CVD) on copper and transferred on our nanowire array by Graphenea. In addition, SLGs on Bi crystal surfaces and Si were fabricated. It is observed that there is strong adhesion after transference. This is not surprising considering van-der-Waals attraction between graphene and bismuth.

Raman spectra were collected at room temperature with the Ranishaw-in Via spectrometer. In Figure 2, we show typical Raman spectroscopy data taken on the SLG sheets after transfer onto the 200 nm bismuth nanowire array. We also studied a SLG coated bismuth (111) single crystal surface. The bismuth crystal sample, which is 1mm thick and has a cross-section of 0.5 cm$^2$, is prepared by cleaving a slab section from a single crystal. The sample is polished mechanically after the cleaving prior to the graphene transfer. For comparison purposes, we also measured a SLG sample on polycrystalline silicon. We also tested a sample of SLG on silicon dioxide. The Raman of SLG on Si that we observe compares favorably with the observations of Tongay *et al*.[49] The Raman lines of graphene on bismuth crystals and on Bi NWA have never been reported before. We observe that the Raman peaks are resolved both in the Bi sample and the Bi nanowire array samples. At the low frequency region, there are peaks at 70 and 100 cm$^{-1}$ which are consistent with the Eg and Ag vibrational modes of Bi.[50] The bismuth peaks indicate that we have a case of physisorption. Raman spectra in the high wavenumber range, that is the 500-3500 cm$^{-1}$ range, are shown in Figure 2. The four characteristic peaks for graphene at 1590 cm$^{-1}$ (G-band), 1350 cm$^{-1}$ (disorder D-band), 2670 cm$^{-1}$ (2D band) and 2900 cm$^{-1}$ (D+D') are observed indicating the unique electronic structure of graphene is preserved. In addition we observe three lines that have not being identified before, namely 1210 cm$^{-1}$ (P1), 2450 cm$^{-1}$ (P2) and 3150 cm$^{-1}$ (P3). The defect D' line at 1600 cm$^{-1}$ is also observable. Our analysis of doping is based on the study of the 2D and G lines that are shown, expanded, in Figure 2b and 2c. The various peaks are fit with Lorentzian functions and the peaks position, amplitudes and full width at half maximum (FWHM) are presented in Table I of the supplement. In comparison to the Raman peaks of graphene on silica, the center frequencies of the G and 2D peaks are shifted by small amounts. It should be noted that the width of the D line is more than twice that of the G and 2D lines. Additionally, to the presence of the D and D' peaks evidences disorder. The disorder that is observed can be characterized following Cançado et al.[37]; they presented a systematic study of the ratio between the integrated intensities of the D and G Raman lines of nanographite with different crystalline sizes. Accordingly, we estimate the crystalline size of our samples to be 20 nm. The SLG disorder that was observed by us can be characterized also using the method of comparison with ion-impact-damaged graphene.[38] Considering that the breath of the D peak is twice that of the G peak, the $I_D/I_G$ ratio of 1.14 is the same that is observed in ion-impact-damaged samples with the mean distance between damage centers $L_D$ of 8 nm. However, here the disorder is associated with the contact of the graphene with the nanowire array and therefore another more sensible interpretation is that the areas directly in contact with the bismuth nanowires become activated because of damage. Experimental and theoretical studies show that an activated area has an $I_D/I_G$ ratio of roughly 4.2. Since the fractional area occupied by Bi nanowires is 30 % this simple model predicts $I_D/I_G \sim 1.2$, which is very close to the experimental value.

Raman can be employed to characterize dopants in graphene and to evaluate the charge density. When SLG is transferred onto a substrate, equilibration of the Fermi level throughout the system gives rise to a charge transfer between the graphene and the substrate, thereby creating the built-in field. The G and 2D bands are both strongly influenced by the carrier concentration and they have been extensively studied for doping characterization.[39-42] These experiments were carried out in undamaged SLG and doping was performed in an electrochemical manner. It was found that the position of the 2D band depends on the Fermi energy $E_F$. The 2D band position increases as the hole concentration increases and decreases as the electron concentration increases. Neglecting the damage in our graphene, our observation of a large 2D line Raman shift of -(45 ± 5) cm$^{-1}$ in graphene on Bi nanowire arrays can be accounted for by an electron doping of approximately (4 ± 0.5) ×10$^{13}$ cm$^{-2}$. This doping level makes the graphene Fermi energy be approximately (0.8 ± 0.08) eV. Such level of doping would also explain the reduced intensity of the 2D line, $I_{2D}$, with respect to the intensity of the G line, $I_G$. The ratio $I_{2D}/I_G$ can be as large as 3.5 for undoped graphene whereas for graphene in contact with bismuth nanowires the ratio of intensities $I_{2D}/I_G$ is only 0.75. The width and position of the G band also change with doping. Khomyakov et al studied the adsorption of graphene on metal substrates using first-principles calculations.[36] It was shown that the bonding of graphene to some metals, notably Cu, with low work function is so weak that the unique graphene electronic structure is preserved. The interaction does, however, lead to a charge transfer, doping, that shifts the Fermi level by up to 0.5 eV with respect to the conical points. Khomyakov *et al* provides insight into our observations. He showed that the difference of work functions of the substrate and SLG is relevant to the graphene doping. A review of the literature shows that the work function of bismuth is between 4.2 and 4.3 eV.[51] The work function of suspended graphene is between 4.9–5.1 eV.[52] According to Khomyakov, because of the work function of bismuth is lower than that of SLG, doping should be is n-type, as observed. Akturk et al have studied the doping of graphene by bismuth atoms using density functional theory and found that Bi can be weakly physisorbed.[53] This result matches our observation that the Raman peaks of Bi are not disturbed. Akturk report charge transfer $Q$ from the Bi atom to graphene to be in the range between 1.5 and 5 electrons per atom. Since the inter-bismuth distance is given as 0.3 nm, assuming a single-particle case, we can estimate the doping to range between $1.5 \times 10^{14}$ and $5 \times 10^{14}$ cm$^{-2}$. Although theory predicts results slightly larger than in the experiment, it is clear that Akturk' work provides a compelling physical picture of the electronic system. Chen *et al* [54] have studied adsorption of Bi atoms and clusters on graphene grown by epitaxy (MEG). They observed that upon Bi deposition, charge transfer from MEG to Bi adatom and a characteristic peak, corresponding to the p-band of Bi, can be observed in the tunneling spectrum.

We studied the photoresponse graphene/nanowire array hybrid devices as well as devices based on indium tin oxide/nanowire array interfaces. In both cases we employed the same equipment consisting of monochromatic light sources consisting of LEDs. The graphene heterostructure device demonstrates broadband photoresponse from ultraviolet (UV) to the near infrared (NIR) range. The response in the UV and visible range is shown in Figure 3a. The peak current is observed at about 580 nm, which may originate from the relatively high power of the light source at this particular wavelength. In the case of devices based on Bi nanowire array/indium tin oxide the photoresponse current (PC) increases with chopping frequency $f$ as PC ~ $f^{1/2}$. This indicates that the signal can be completely described by thermoelectric effects considering cooling rates given by heat diffusion. In contrast, in the graphene bismuth samples, a completely different frequency dependence is observed. This is presented in Figure 3b. PC increases with decreasing chopping frequency and the trend that is uniformly observed for all the wavenumbers is PC ~ $f^{-0.9}$. We observe this power-law frequency dependence of the photoresponse over three order of magnitude of excitation frequency consistent with a wide distribution of diffusion times. A very interesting model of transport was proposed by Nokidov *et al*[55] to explain power-law current transients observed in partially ordered arrays of semiconducting nanocrystals. The model describes electron transport by a stationary Lévy process of transmission events and thereby requires no time dependence of system properties.

      Figure 4 presents the energy diagram of the interface of the Bi-NWA-SLG hybrid. This diagram is similar to the one for metals. Work functions $W_M$ of bismuth and $W_G$ of graphene are represented. Also, the Fermi energy $E_F$ of the electrons (metal and graphene) is represented. The screening length is represented by $\lambda$. $\Delta E_F$ ($\Delta E_F < 0$ because SLG is n-doped) is the difference of the graphene Fermi energy with respect to the Dirac point. We estimate, on the basis of Raman spectroscopy results, that $\Delta E_F \sim$ −0.8 eV. We illustrate the interface dipole, the potential step formation, where $\Delta V$ is the voltage drop produced over the contact, and the electric field due to the dipole. The actual work function of graphene ($W = W_G + \Delta E_F$). There is no rectifying Schottky barrier but rather an Ohmic contact as observed in our studies of the I-V of the interface.

CONCLUSION

In summary we have demonstrated a new room-temperature photodetector based on semimetal bismuth nanowire arrays coupled with a graphene substrate. Our heterostructure addresses the major challenge that is the synthesis of well-defined low-dimensional structures of uniform nanosizes by combining nanowires with atomic layers. Unlike detectors based on the conductivity of bismuth thin films or nanowires, our photodetector principle is the current generated at the heterostructure built-in fields. While the structural flatness and high electron mobility of graphene exhibit great potential for future generations of electronic materials, its high transparency imposes severe limits on applications for photodetection. This, coupled with bulk bismuth high reflectance, is a shortcoming that our novel architecture addresses. Nanowire arrays coupled with graphene have strong absorption because of light trapping and other metamaterial effects. In addition, bismuth, as a semimetal, complements graphene's gapless and mobility characteristics and features a long screening length (20 nm) comparable to semiconductors. This property dramatically enhances optical absorption. We apply Raman spectroscopy to evaluate the quality of the interface, probe the interfacial doping, analyze the electrical properties of the bismuth graphene junctions in terms of contact resistance (there is no evidence of a Schottky barrier), and characterize the charge transfer between graphene and Bi. The analyzed spectrum ranges from the visible (350 nm) through the near infrared (980 nm) and well into the telecommunication band (1550 nm). Because of photocarrier pair generation and transfer at the interface, the photocurrent generated by the interface built-in field is robust without sacrificing the detection spectrum. Our results for the interface energies and charge compare favorably with those recently published for Bi-adatoms on graphene. Intriguingly, for the photoresponse we observe a power-law frequency dependence over three order of magnitude of excitation frequency consistent with a wide distribution of diffusion times. This characterization fits the model of Levy paths that explains anomalous transport in quantum-dot arrays. This is the first instance that fractional statistics have been proposed for photocurrent generation. These key properties may enable application of junction of graphene with Bi devices in detector and light harvesting applications. Owing to the built-in fields, no direct bias voltage between the bismuth and graphene is needed to insure photocurrent generation enabling new applications such as solar energy harvesting.

METHODS

The alumina template used in this work was sold commercially under the trade name Anopore (Whatman, MA) and was made by the anodization of aluminum. The template is a 55 microns thick plate of alumina which supports an array of parallel cylindrical channels running perpendicular to the plate. An illustration of the nanowire array fabrication process is presented in the Supplement. The alumina plates and the bismuth were packed in a thin glass tube. The tube was closed at the bottom. The tube was then inserted into the reactor of a high temperature/high pressure injection apparatus. The molten bismuth was injected in the channels of the alumina template. Pressure was needed to overcome the effect of the surface tension combined with the non-wetting characteristics of bismuth on the template walls. A reactor and pump rated for 10 kBar service were employed. The reactor temperature is 300 C. When the injection was complete, the melt solidified inside the alumina channels. The single layer graphene film (SLG) was fabricated by chemical vapor deposition (CVD) on copper, detached and transferred on our nanowire array by Graphenea. In addition, SLGs on Bi crystal surfaces and Si were fabricated. We also studied a SLG coated bismuth (111) single crystal surface. The bismuth crystal sample, which is 1mm thick and has a cross-section of 0.5 $cm^2$, is prepared by cleaving a slab section from a single crystal. The sample is polished mechanically after cleaving prior to the graphene transfer. For comparison purposes, we also measured a SLG sample on polycrystalline silicon. This sample was purchased from ACS, Medford, MA, USA. We also tested a sample of SLG on silicon dioxide (Graphenea, San Sebastián, Spain). Raman spectra were collected at room temperature with the Ranishaw-in Via spectrometer. The laser wavelength is 514.5 nm and the laser power is 0.5 mW. The area of the laser spot in smaller than 1 $\mu m^2$ and the spectral resolution is better that 1 $cm^{-1}$. The surface was observed by scanning electron microscopy.

*Conflict of interest*: The author declare no competing financial interest.

*Acknowledgement*: This work was supported by the National Science Foundation through Partnerships for Research and Education in Materials (PREM), effort PRDM 1205608 that involves Howard University and Cornell University. The National Science Foundation also supported the Science and Technology STC 1231319. This STC is denominated Center for Integrated Quantum Materials (CIQM) and involves Harvard University, Howard University and MIT. We also acknowledge support by The Boeing Company.

*Supporting Information Available*: Additional experimental details and figures. This material is available free of charge via the internet at http://pubs.acs.org.

FIGURES

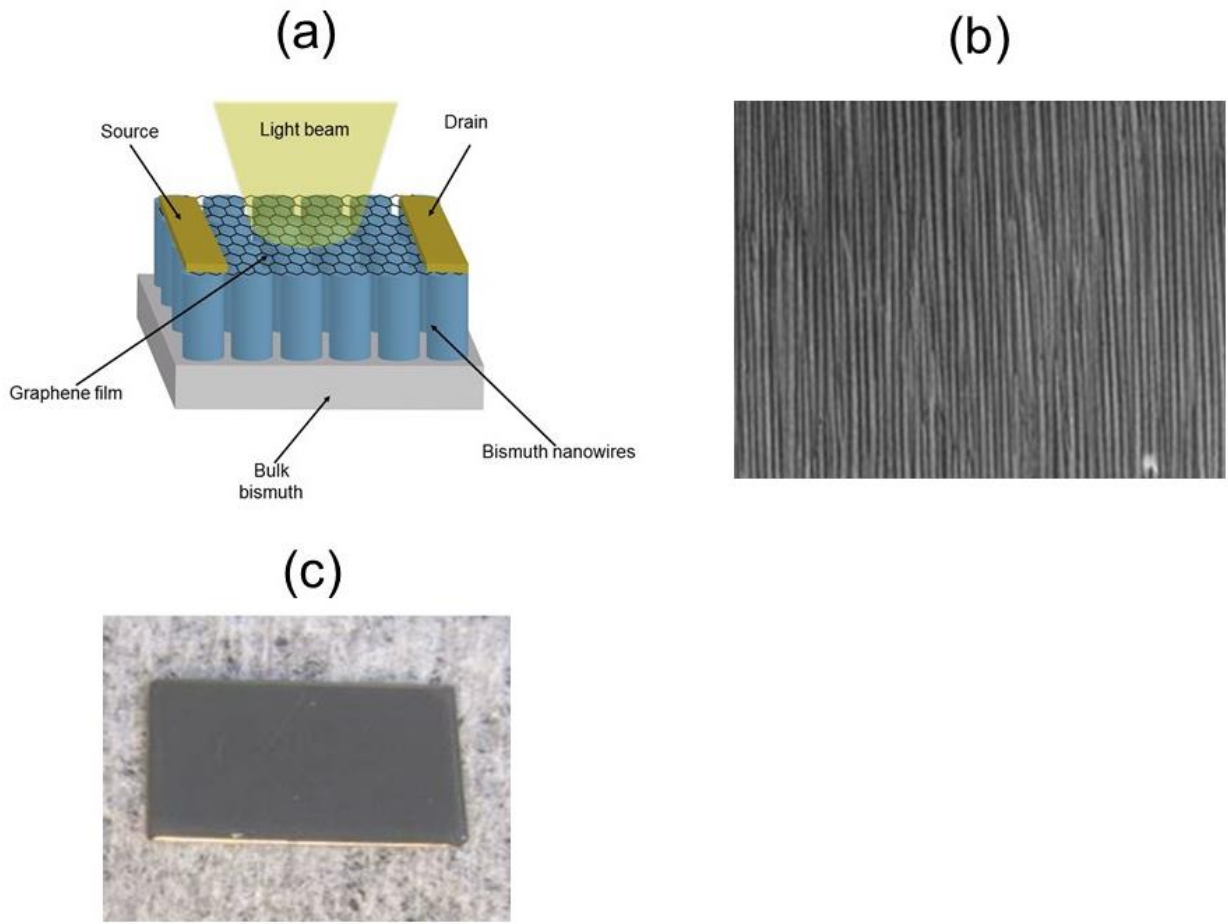

**Figure 1. Bi Nanowire Array Array Device Structure. (a) Schematic design of the photodetector device. The I(V) between the bulk bismuth base and the drain is linear and the resistance is 90 Ohms. (b) SEM image of the nanowire array. (c) Optical microscope image of the nanowire array showing that the nanowire array appears dark.**

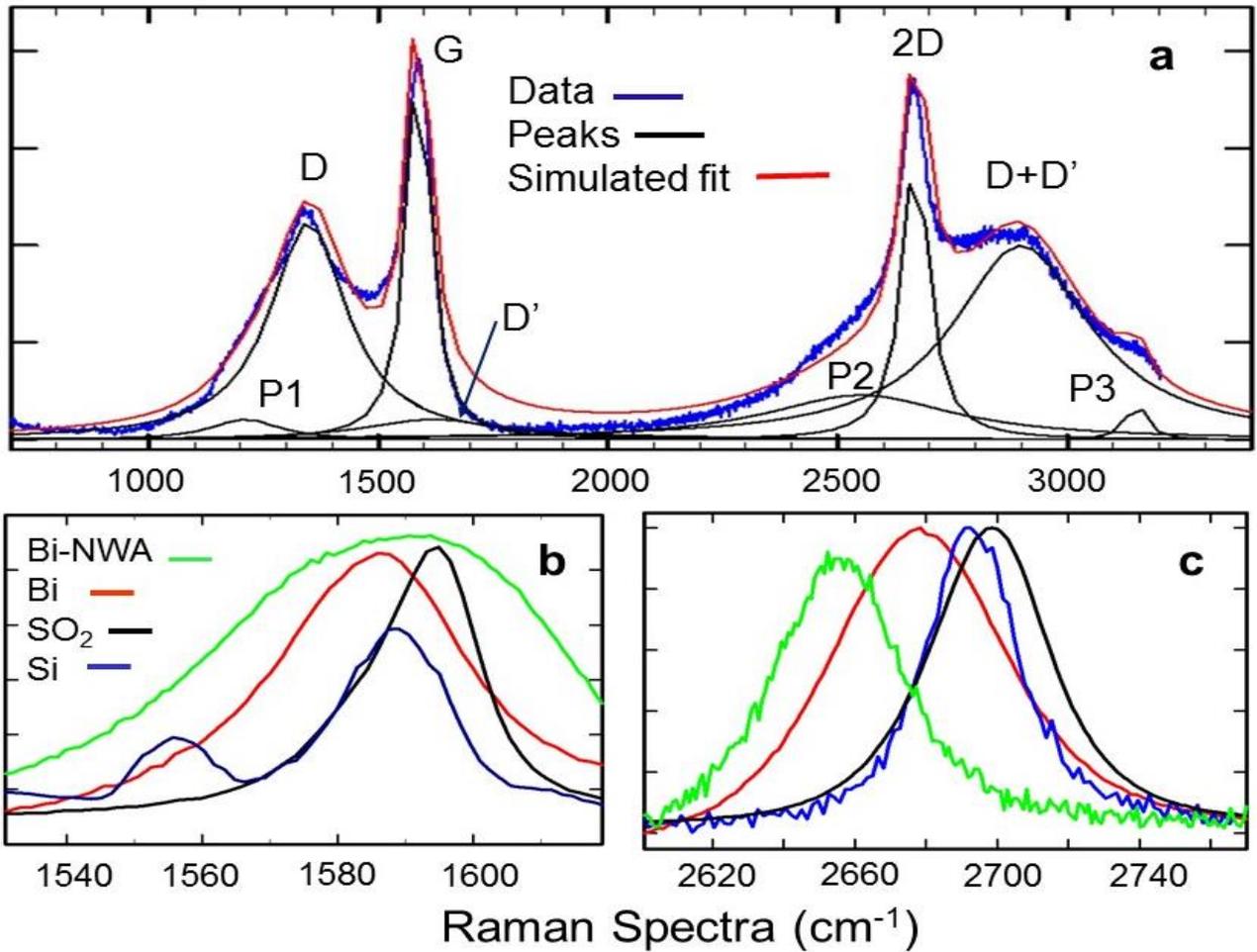

**Figure 2. Raman spectra of Bi Nanowire Array Graphene Heterostructure.** We show the data after subtraction of a smooth background, the individual peaks and the simulated fit. (a) CVD-grown graphene on Cu foils after transfer onto various substrates. (b) The Raman spectra of the G peak. The black curves are the measurements on graphene/SiO$_2$, that is our reference for **shifts, and the other curves are for the graphene surface combinations as indicated in the legend.**

**(c) Same as in (b) for the Raman spectra of the 2D peak.**

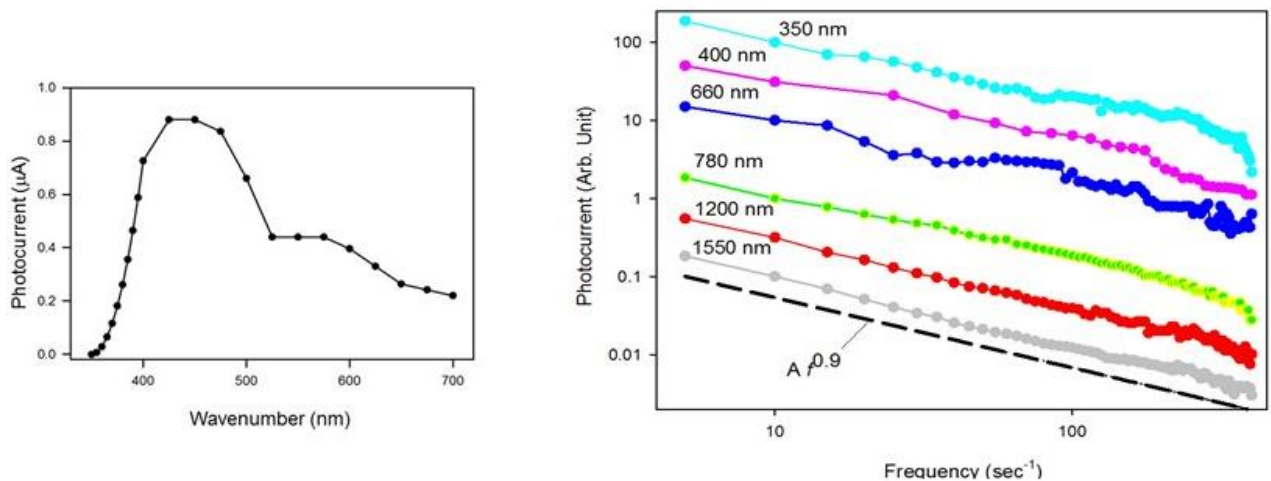

**Figure 3.** (a) Photocurrent of the heterostructure device as a function of the photoexcitation wavelength (from 300 nm to 700 nm). The photocurrent is measured between the bulk bismuth base and the drain (Figure 1) that is in contact with the graphene. (b) Detector Intensities versus chopping frequency $f$ for various wavenumbers. The dashed line represents a power-law dependence of the PC with an exponent of 0.9.

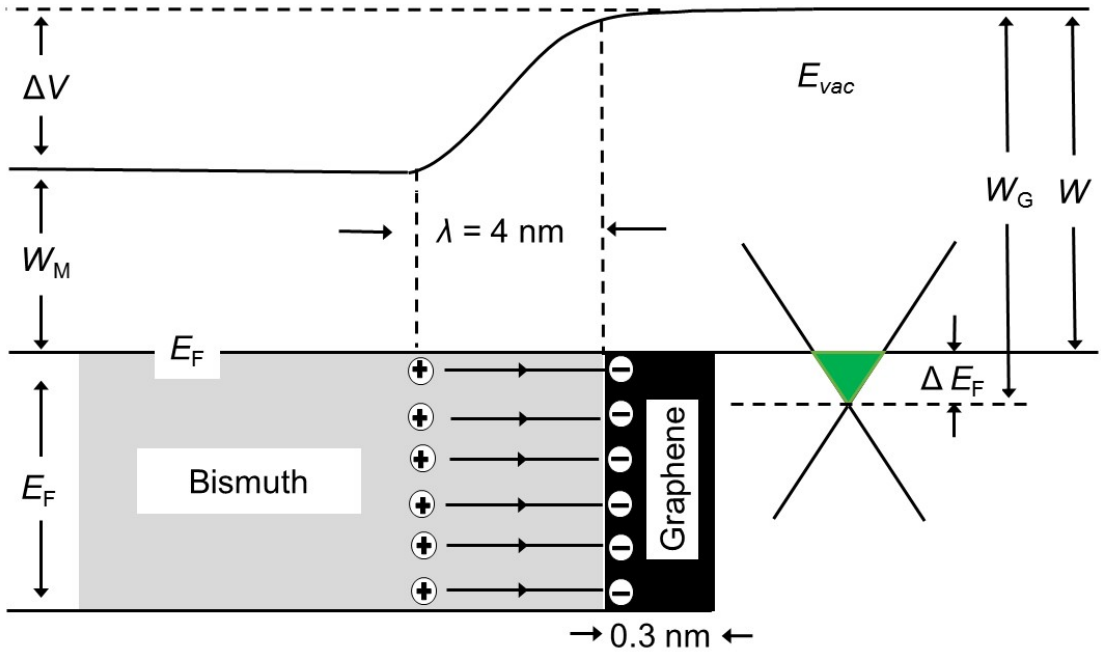

**Figure 4. Carrier transport processes at the bismuth-graphene junction.** Schematic view of the band profile and dipole formation at the semimetal-graphene interface. Work functions $W_M$ of bismuth and $W_G$ of graphene are represented. Also, the Fermi energy $E_F$ of the electrons (metal and graphene) is represented. The screening length is represented by $\lambda$. is the shift of neutrality point of graphene due to the metal doping is $\Delta E_F$ ($\Delta E_F \sim -0.8$ eV). We illustrate the interface dipole and the potential step $\Delta V$ that is the voltage drop produced over the contact and the electric field due to the dipole. The actual work function of graphene $W$ is also shown.

Supplementary Material

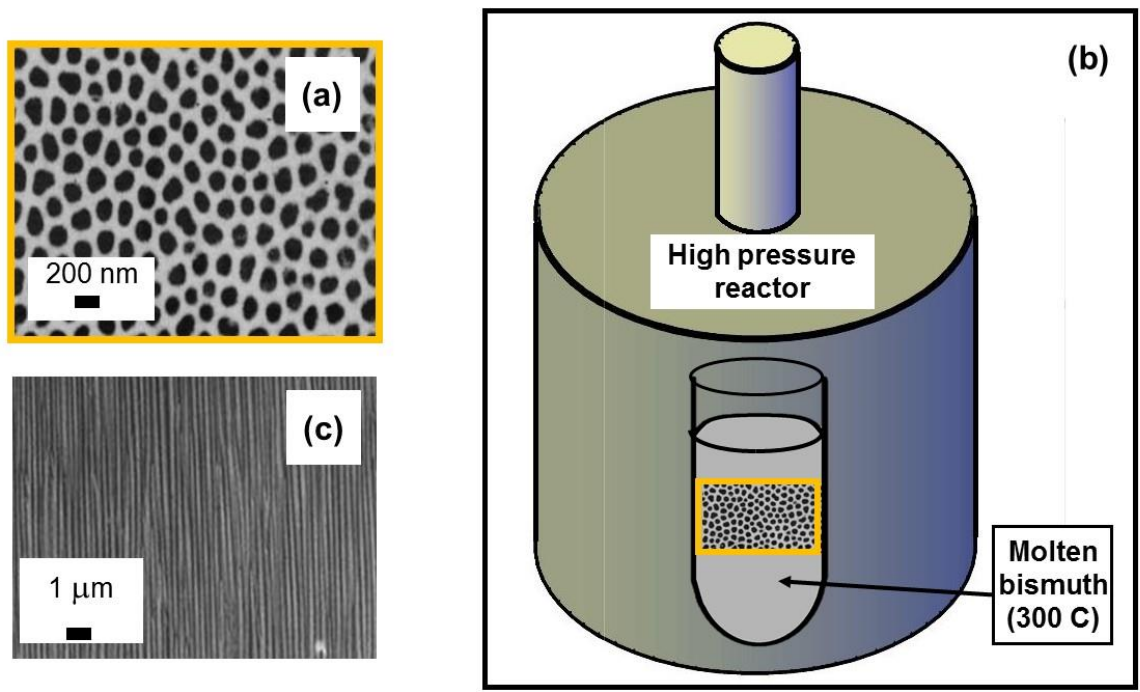

**Supplementary Figure 1. Fabrication of Bi nanowire array using the high-pressure injection method. (a) Porous alumina template, top view. Dark and light areas represent nanochannels and alumina, respectively. (b) High pressure reactor with the glass tube with a closed bottom containing the template in molten bismuth. After injection, bismuth embeds the nanochannels. (c) SEM image of the nanowire array, side view. It shows the nanowires that are formed in the nanochannels. The alumina is not very visible due to high contrast between the alumina insulator and the Bi nanowires.**

|       | Position (cm$^{-1}$) | Intensity | HWHM(cm$^{-1}$) |
|-------|---------------------|-----------|-----------------|
| P1    | 1210                | 0.056     | 210             |
| D     | 1340                | 0.62      | 220             |
| G     | 1580                | 1         | 70              |
| D'    | 1620                | 0.04      | 370             |
| P2    | 2530                | 0.13      | 550             |
| 2D    | 2660                | 0.75      | 80              |
| D+D'  | 2900                | 0.57      | 360             |
| P3    | 3150                | 0.09      | 80              |

**Table I. The various Raman peaks that were observed, see in figure 2, are fit with Lorentzian functions and the table presents the peaks position, amplitudes relative to the G peak and full width at half maximum (FWHM).**

Screening Length